\documentclass{IEEEtran}
\usepackage{url}
\usepackage{amsmath}
\usepackage{amssymb}
\usepackage{amsthm}
\usepackage{authblk}
\usepackage{upgreek}
\usepackage[ruled,algonl,lined]{algorithm2e}
\usepackage{cite}
\usepackage[dvipsnames]{xcolor}
\usepackage{comment}
\usepackage{hyperref}

\newtheorem*{corollary*}{Corollary}

\usepackage[normalem]{ulem}

\DeclareMathAlphabet{\mathbit}{OML}{cmr}{bx}{it}
\DeclareMathAlphabet{\mathsf}{OT1}{cmss}{m}{n}
\DeclareMathAlphabet{\mathbsf}{OT1}{cmss}{bx}{it}

\DeclareMathOperator{\sign}{sign}

\newcommand{\set}[1]{{\cal #1}}

\newcommand{\kt}{\tilde{k}}

\newcommand\diff[1]{\ensuremath{\:\mathrm{d}#1}}

\newcommand\pderiv[2]{\ensuremath{\frac{\partial#1}{\partial#2}}}
\newcommand\pderivk[3]{\ensuremath{\frac{\partial^{#3}#1}{\partial{#2}^{#3}}}}
\newcommand{\Deltarm}{\ensuremath{\mathrm{\Delta}}}

\DeclareMathOperator{\sinc}{sinc}

\newlength{\figurewidth}
\newlength{\figureheight}

\usepackage{tikz}
\usetikzlibrary{arrows,decorations,backgrounds,shadows,plotmarks,calc, positioning}
\usetikzlibrary{arrows.meta}

\usepackage{pgfplots}
\pgfplotsset{compat=newest}
\pgfplotsset{plot coordinates/math parser=false}
\pgfplotsset{every axis/.append style={font=\footnotesize}}
\pgfplotsset{
	ylabel right/.style={
		after end axis/.append code={
			\node [rotate=90, anchor=north] at (rel axis cs:1,0.5) {#1};
		}   
	}
}

\usepackage{fancyhdr}
\fancypagestyle{first}{%
	\chead{\footnotesize This is the author's version of an article that has been published in this journal. Changes were made to this version by the publisher prior to publication.\\
		The final version of record is available at
		\href{http://dx.doi.org/10.1109/JLT.2020.3021277}{http://dx.doi.org/10.1109/JLT.2020.3021277}
	}
	\rhead{\thepage}
	\cfoot{\scriptsize Copyright (c) 2020 IEEE. Personal use is permitted. For any other purposes, permission must be obtained from the IEEE by emailing pubs-permissions@ieee.org.
	}
}

\fancypagestyle{others}{%
	\rhead{\thepage}
	\fancyfoot{}
}

\title{Mismatched Models to Lower Bound the Capacity of Optical Fiber Channels}
\author{Francisco Javier Garc\'ia-G\'omez and Gerhard Kramer \IEEEmembership{Fellow, IEEE}
\IEEEcompsocitemizethanks{
	\IEEEcompsocthanksitem Date of current version August 27, 2020. J. Garc\'ia-G\'omez (e-mail: javier.garcia@tum.de) and Gerhard Kramer are with the Institute for Communications Engineering (LNT), Technical University of Munich, Germany. This work was supported by the German Research Foundation (DFG) under Grants KR 3517/8-1 and 3517/8-2.
	}
	}  

\begin{document}
\maketitle

\begin{abstract}
A correlated phase-and-additive-noise (CPAN) mismatched model is developed for wavelength division multiplexing over optical fiber channels governed by the nonlinear Schr\"odinger equation. Both the phase and additive noise processes of the CPAN model are Gauss-Markov whereas previous work uses Wiener phase noise and white additive noise. Second order statistics are derived and lower bounds on the capacity are computed by simulations. The CPAN model characterizes nonlinearities better than existing models in the sense that it achieves better information rates. For example, the model gains 0.35 dB in power at the peak data rate when using a single carrier per wavelength. For multiple carriers per wavelength, the model combined with frequency-dependent power allocation gains 0.14 bits/s/Hz in rate and 0.8 dB in power at the peak data rate.
\end{abstract}


\thispagestyle{first}
\pagestyle{first}

\section{Introduction}
Computing the capacity of nonlinear optical channels is an open problem. The spectral efficiency of the additive white Gaussian noise (AWGN) channel, $\log_2(1+\text{SNR})$ where $\text{SNR}$ is the signal-to-noise ratio, is known to be an \emph{upper} bound on the spectral efficiency of optical channels modeled by the nonlinear Schr\"odinger equation (NLSE)~\cite{kramer2015upper,yousefi2015upper}. All existing \emph{lower} bounds reach an information rate peak at some launch power and then decrease as the launch power increases. Over the years, many simplified models of the NLSE have been developed to obtain better lower bounds, see the review in~\cite[Sec.~I.A]{ghozlan}. Our focus is on the regular perturbation (RP) approach that was developed for dispersion-compensated on-off keying systems with Gaussian pulses in~\cite{mecozzi2000analysis, mecozzi2000system, mecozzi2001cancellation} and that was formalized for arbitrary waveforms in~\cite{vannucci2002rp}.

Numerical lower bounds on the capacity were developed in~\cite{essiambre_limits} by using wavelength division multiplexing (WDM), geometric and probabilistic shaping, and by treating nonlinear distortions as additive Gaussian noise (AGN) whose statistics depend on the transmit symbol amplitude (non-Gaussian noise was also tested~\cite[Sec.~X.C]{essiambre_limits}). This model was refined in~\cite{poggiolini2014gn, carena2014egn}. In~\cite{mecozzi2012nonlinear, dar2013properties, dar2014new}, regular perturbation (RP) led to models with correlated phase noise, improving the lower bounds in~\cite{essiambre_limits}. In~\cite{secondini2012analytical, secondini2013achievable}, logarithmic perturbation (LP) resulted in a time-variant phase noise model, see also~\cite{frey2019discrete}. This model motivates using multi-carrier modulation and particle filtering~\cite{secondini2017fiber, secondini2019nonlinearity} which achieves the best lower bounds that we are aware of.

We propose an extension of the RP and LP models to include memory in both the phase and additive noise processes. We also replace the Wiener phase noise with a Gauss-Markov process. This change does not increase the information rates, but it better fits the statistics of the NLSE. On the other hand, including memory in the additive noise improves the best capacity lower bounds known to us~\cite{secondini2017fiber}. For example, the model gains 0.35 dB in power efficiency at the information rate peak with a single carrier per wavelength. We further show that a frequency-dependent power allocation for multi-carrier systems gains 0.14 bits/s/Hz over the best existing rate bounds and 0.8 dB in power at the peak data rate.

We remark that a combined regular and logarithmic perturbation (CRLP) analysis of the NLSE was proposed in~\cite{secondini2009crlp} where the focus is on single-channel transmission and self-phase modulation. Most capacity studies focus on WDM where cross-phase modulation (XPM) is the limiting factor.

\subsection{Notation}
Random variables are written with uppercase letters such as $X$ and their realizations with the corresponding lowercase letters $x$. The expectation of $X$ is denoted by $\langle X \rangle$.
The expectation of $X$ conditioned on $Y=y$ is written as $\langle X | Y=y \rangle$. The inner product of the signals $a(t)$ and $b(t)$ with time parameter $t$ is written as $\langle a(t),b(t) \rangle$. The mutual information of two random variables $X$ and $Y$ with joint density $p(x,y)=p(x) p(y|x)$ is given by
\begin{align}
I(X;Y) %
& =\left\langle \log_2 \frac{p(Y|X)}{p(Y)}\right\rangle
= h(Y) - h(Y|X)
\label{eq:mutual-information}
\end{align}
where the entropies are
\begin{align}
& h(Y) = \left\langle - \log_2 p(Y) \right\rangle \label{eq:entropy} \\
& h(Y|X) = \left\langle - \log_2 p(Y|X) \right\rangle.
\label{eq:conditional-entropy}
\end{align}
We write the density of $X$ as $p_X(\cdot)$ if the argument is not the lower-case version of the random variable.

Vectors are denoted as $\mathbf{x}=(x_1,\dots,x_L)$ and signals as $x(t)$. The Euclidean norms of $\mathbf{x}$ and $x(t)$ are written as $\|\mathbf{x}\|$  and $\| x(t) \|$, respectively. The Fourier transform of $u(t)$ is $\set{F}(u(t))(\Omega)$, where $\Omega$ denotes angular frequency. The inverse Fourier transform of $U(\Omega)$ is $\set{F}^{-1}(U(\Omega))(t)$. The dispersion operator $\mathcal{D}_z$ is described by
\begin{align}
   \mathcal{D}_z u(t) = \mathcal{F}^{-1}\left( e^{j \frac{\beta_2}{2} \Omega^2 z} \mathcal{F}(u(t)) \right) .
\end{align}
We write $\sinc(t) = \frac{\sin(\pi t)}{\pi t}$. We write $\delta[\ell]$ for the function that maps integers to zero except for $\delta[0]=1$.

We consider the following parameters.
The variables $z$ and $t$ represent distance and time, respectively. The symbol period is $T$. The launch position is $z=0$ and the receiver position is $z=\mathcal{L}$. The fiber span length is $L_s$. The attenuation coefficient is $\alpha$, the dispersion coefficient is $\beta_2$, and the nonlinearity coefficient is $\gamma$.

\section{Nonlinear Schr\"odinger Equation and RP}

The NLSE~\cite{agrawal_nfo} describes propagation along an optical fiber:
\begin{equation}
\pderiv{}{z}u=-j\frac{\beta_2}{2}\pderivk{}{t}{2}u+j\gamma f(z)|u|^2u+\frac{n(z, t)}{\sqrt{f(z)}}
\label{eq:nlse_u}
\end{equation}
where $u(z, t)$ is the propagating signal and $f(z)$ models loss and amplification along the fiber~\cite{mecozzi2012nonlinear}. We have $f(z)=1$ for ideal distributed amplification (IDA) and $f(z)=\exp(-\alpha z+\alpha L_s\left\lfloor z/L_s\right\rfloor)$ for lumped amplification. For a receiver with bandwidth $\mathcal{B}$, the accumulated noise at $z=\mathcal{L}$ is usually dominated by amplified spontaneous emission (ASE) noise  with autocorrelation function
$N_{\textrm{ASE}} \mathcal{B}\, \sinc(\mathcal{B}(t-t'))$.

\subsection{Continuous-Time RP model}

RP~\cite[Sec.~III]{mecozzi2012nonlinear} expands $u(z, t)$ in powers of $\gamma$
\begin{equation}
u(z, t)=u_0(z, t)+\gamma\Deltarm u(z, t)+\mathcal{O}(\gamma^2).
\label{eq:u_rp}
\end{equation}
Assuming $\gamma$ is small, the right-hand side of~\eqref{eq:u_rp} is placed in~\eqref{eq:nlse_u}, and the equations are solved for the zeroth and first powers of $\gamma$. The result is
\begin{equation}
u(z, t)=u_0(z, t)+u_{\mathrm{NL}}(z, t)+\mathcal{O}(\gamma^2)
\label{eq:rp_full}
\end{equation}
where the linear terms are
\begin{align}
& u_0(z, t)=\mathcal{D}_z \left[ u(0, t) + u_{\mathrm{ASE}}(z, t) \right] \\
& u_{\mathrm{ASE}}(z, t)=\int_{0}^{z}\mathcal{D}_{-z'}\left(\frac{n(z', t)}{\sqrt{f(z')}}\right)\diff z'
\label{eq:u_ASE}
\end{align}
and the \textit{nonlinear perturbation} term is
\begin{equation}
u_{\mathrm{NL}}(z, t)=j\gamma \mathcal{D}_z\left[\int_{0}^{z} f(z')\mathcal{D}_{-z'}\left[|u_0(z', t)|^2 u_0(z', t)\right]\diff z'\right].
\label{eq:u_NL}
\end{equation}

The nonlinear term is responsible for signal-signal mixing and signal-noise mixing because $u_0(z,t)$ includes noise. We focus on WDM systems where the limiting factor is XPM, so we ignore signal-noise mixing and replace $u_0(z', t)$ by $\mathcal{D}_{z'} u(0, t)$ in~\eqref{eq:u_NL}.

\subsection{Discrete-Time RP Model}
Consider WDM and pulse amplitude modulation (PAM) with $C$ channels with indexes $c$ between $c_{\textrm{min}}\le 0$ and $c_{\textrm{max}}\ge 0$:
\begin{align}
   c\in\set{C}=\left\{c_\textrm{min},\ldots,-1,0,1,\ldots,c_\textrm{max}\right\}
\end{align}
with $C=c_{\textrm{max}}-c_{\textrm{min}}+1$. Let the channel of interest have index $c=0$. The launch signal including all channels is
\begin{align}
   u(0,t) = \sum_m x_m s(t-mT) + \sum_{c\ne 0} e^{j \Omega_c t} \sum_k b_{c,k} s(t- kT)
\end{align}
where $\Omega_c$ is the center frequency of channel $c\in\set{C}$, and $\Omega_0=0$. The pulse shaping
filter $s(t)$ is taken to be a normalized root-Nyquist pulse with $\| s(t)\|=1$ and
$\langle s(t),s(t+nT) \rangle = 0$ for $n \ne 0$. 

For the modulation, we model the sequences $\{X_m\}$ and $\{B_{c,m}\}$, $c \in \set{C}\setminus\{0\}$,
as being independent, and as each having independent and identically distributed (i.i.d.) and proper complex symbols
with energies $\langle |X_m|^2 \rangle = E$ and $\langle |B_{c,m}|^2 \rangle = E_c$ for all $m$ and $c$. Note that the optical power of channel $c$ is $\mathcal{P}_c=E_c/T$. The properness ensures that the pseudo-covariances are zero.
Since we will need fourth moments, we define $\langle |B_{c,m}|^4 \rangle = Q_c$ for all
$m$ and $c$.

After digital back-propagation of the center channel, and matched filtering and sampling, the discrete-time mismatched model based on RP is (see~\cite[Sec.~VI-VIII]{mecozzi2012nonlinear})
\begin{align}
    y_m = x_m + w_m + \sum_{c\ne 0} \Delta x_{c,m}
    \label{eq:RP}
\end{align}
where the noise realization is
\begin{align}
   w_m = \langle u_{\textrm{ASE}}(t), s(t-mT) \rangle
\end{align}
and the noise process $\{W_m\}$ is i.i.d., circularly-symmetric, complex Gaussian with variance $\sigma_W^2=N_{\textrm{ASE}}$,
and $\{W_m\}$ is independent of $\left\{X_m\right\}$ and $\{B_{c, m}\}$ for all $c$. The non-linear interference (NLI) terms are (see~\cite[Eq.~(60)]{mecozzi2012nonlinear})
\begin{align}
   \Delta x_{c,m} = j \sum_{n,k,k'} C_{n,k,k'}^{(c)} \cdot x_{n+m} \cdot b_{c,k+m}\, b_{c,k'+m}^*
   \label{eq:NLI-term}
\end{align}
where the NLI coefficients are
\begin{align}
    & C_{n,k,k'}^{(c)} =
    2\gamma \int_0^{\mathcal{L}} f(z) \left[ \int_{-\infty}^{\infty} s(z,t)^*\, s(z,t-nT) \right. \nonumber \\
    & \quad s(z,t - kT + \beta_2 \Omega_c z)\, s(z,t - k'T + \beta_2 \Omega_c z)^*\, dt \Big] dz
    \label{eq:C_nkk}
\end{align}
and $s(z,t)=\mathcal{D}_z s(t)$. Note that $s(z,t)$ is in general complex-valued even if the
pulse $s(t)$ is not. Observe also that
\begin{align}
C_{n, k, k'}^{(c)}=\left(C_{-n, k'-n, k-n}^{(c)}\right)^* . \label{eq:C1} 
\end{align}
In particular, we have $C_{0,k,k'}^{(c)}=(C_{0,k',k}^{(c)})^*$ and $C_{0,k,k}^{(c)}$ is real.

\begin{figure}[tbp]
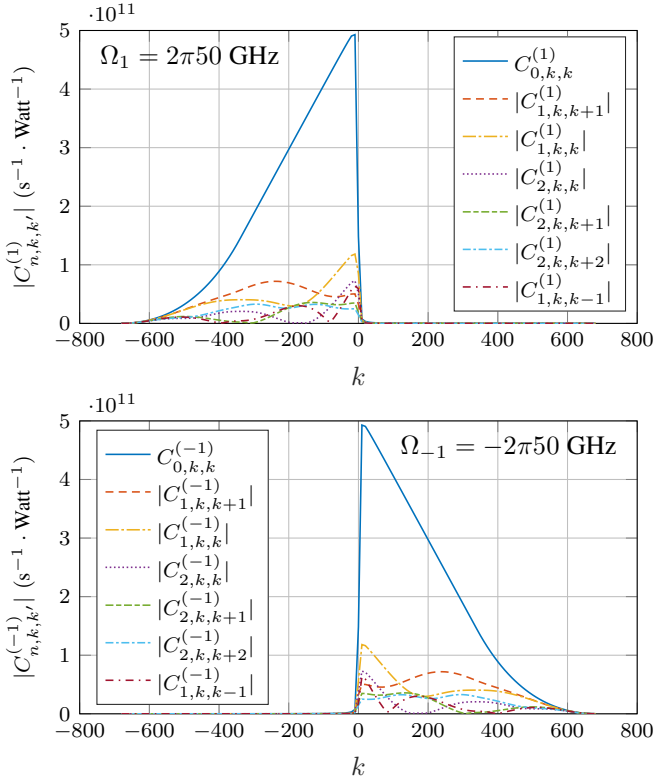
\centering
	\setlength{\figurewidth}{0.88\linewidth}
	\setlength{\figureheight}{0.5\figurewidth}
	\input{Graphs/Cnkk.tex}
	\input{Graphs/Cnkk_minus.tex}
	\caption{NLI coefficients $C_{n, k, k'}^{(c)}$ with $\Omega_1=2\pi 50$ GHz and $\Omega_{-1}=-2\pi 50$ GHz for a $1000$-km link with the parameters in Table~\ref{tab:parameters}. Note that $C_{0, k, k+1}^{(c)}=C_{1, k+1, k+1}^{(c)}$ and $C_{0, k, k+2}^{(c)}=C_{2, k+2, k+2}^{(c)}$.}
	\label{fig:C_nkk}
\end{figure}

Several NLI coefficient magnitudes $|C_{n, k, k'}^{(1)}|$ are plotted in the top of Fig.~\ref{fig:C_nkk} with $\Omega_1=2\pi 50$ GHz. Observe that $C_{0, k, k}^{(1)}$ has the largest magnitude. The top of Fig.~\ref{fig:C_nkk} also shows $|C_{n, k, k'}^{(1)}|$ for $0<k\le 680$, but these NLI coefficients are very small. Due to different group velocities, the symbols from the channel $\Omega_1=2\pi 50$ GHz that interfere with the channel of interest are mostly past symbols ($k\le 0$). The bottom of Fig.~\ref{fig:C_nkk} shows $|C_{n, k, k'}^{(-1)}|$ for the channel $\Omega_{-1}=-2\pi 50$. Note that if $\Omega_c=-\Omega_{-c}$, then we have
\begin{equation}
    C_{n, k, k'}^{(-c)} = C_{-n, -k, -k'}^{(c)} = (C_{n, n-k', n-k}^{(c)})^* \label{eq:C3} 
\end{equation}
where the last step is the same as~\eqref{eq:C1}. For example, the curve $|C_{0,k,k}^{(-1)}|$ is the same as $|C_{0,k,k}^{(1)}|$ but flipped at position $k=0$. Similarly, the curve $|C_{1,k,k}^{(-1)}|$ is the same as the curve $|C_{1,k,k}^{(1)}|$ but flipped at position $k=1$ of the top plot.

\section{CPAN Mismatched Model}
\label{sec:cpan}

We separately consider the NLI involving $x_m$ and $\{x_n\}_{n\ne m}$. Consider \eqref{eq:NLI-term} and collect the terms as
\begin{align}
   \Delta x_{c,m}
   & = j x_m \theta_{c,m} + v_{c,m}
   \label{eq:Delta-expand}
\end{align}
where
\begin{align}
   \theta_{c,m} & := \sum_{k,k'} C_{0,k,k'}^{(c)} \cdot b_{c,k+m}\, b_{c,k'+m}^* \label{eq:theta} \\
   v_{c,m} & := j \sum_{n\ne 0,k,k'} C_{n,k,k'}^{(c)} \cdot x_{n+m} \cdot b_{c,k+m}\, b_{c,k'+m}^* . \label{eq:nu}
\end{align}
We now approximate
\begin{align}
   x_m (1 + j \theta_{c,m}) \approx x_m e^{j \theta_{c,m}}
   \label{eq:e-approx}
\end{align}
i.e., we view $\theta_{c,m}$ and $v_{c,m}$ as representing phase noise and additive noise, respectively. Our new mismatched model is thus a \emph{correlated phase-and-additive-noise}
(CPAN) model
\begin{align}
    y_m = x_m e^{j \theta_{m}} + w_m + v_{m}
    \label{eq:CPAN}
\end{align}
where
\begin{align}
    \theta_m = \sum_{c\ne 0} \theta_{c,m} \text{ and } 
    v_m = \sum_{c\ne 0} v_{c,m}.
\end{align}
This model is the same as \eqref{eq:RP} except for the change
\eqref{eq:e-approx}. We are thus considering a mixed RP and LP model.

Finally, we remark that $\theta_{c,m}$ is real-valued, because
\begin{align}
   \theta_{c,m}^* & = \sum_{k,k'} (C_{0,k,k'}^{(c)})^* \cdot b_{c,k+m}^*\, b_{c,k'+m} \\
   & = \sum_{k',k} C_{0,k',k}^{(c)} \cdot b_{c,k'+m}\, b_{c,k+m}^* = \theta_{c,m}
\end{align}
where we have used $C_{0,k,k'}^{(c)}=(C_{0,k',k}^{(c)})^*$.

\subsection{First- and Second-Order Statistics}
\label{sec:stat2}
We are interested in studying the first- and second-order statistics of
the symbol sequences $\{X_m\}$ and $\{\Theta_{c,m}\}$, $\{V_{c,m}\}$.

\subsubsection{Means}
The mean values are:
\begin{align}
   & \langle X_m \rangle = \langle V_{c,m} \rangle = 0 \\
   & \langle \Theta_{c,m} \rangle
   = \sum_{k,k'} C_{0,k,k'}^{(c)} \langle B_{c,k+m}\, B_{c,k'+m}^* \rangle = \sum_{k} C_{0,k,k}^{(c)}\, E_c . \label{eq:Theta-mean}
\end{align}
If we condition on $\{X_\ell\}=\{x_\ell\}$, then we have
\begin{align}
   \langle V_{c,m} | \{X_\ell\}=\{x_\ell\} \rangle
   & = \sum_{n\ne 0} \left( j \sum_{k} C_{n,k,k}^{(c)} E_c \right) \cdot x_{n+m}
\end{align}
and one can view this term as an intersymbol interference (ISI).
The complete ISI at time $m$ is therefore
\begin{align}
    \sum_{n\ne 0} \left( j \sum_{c \ne 0}  \sum_{k} C_{n,k,k}^{(c)} E_c \right) \cdot x_{n+m}.
\end{align}

\subsubsection{Second-Order Statistics for $\Theta_{c,m}$}
Consider the (real-valued) phase noise $\{\Theta_{c,m}\}$.

The covariances are %
\begin{align}
   r_{\Theta}^{(c)}[\ell] & := \langle \Theta_{c,m} \Theta_{c,m+\ell} \rangle
   - \langle \Theta_{c,m} \rangle \langle \Theta_{c,m+\ell} \rangle \nonumber \\
   & = \sum_{k} C_{0,k,k}^{(c)} C_{0,k-\ell,k-\ell}^{(c)} (Q_c - E_c^2) \nonumber \\
   & \quad + \sum_{k, k' \ne k} C_{0,k,k'}^{(c)} (C_{0,k-\ell,k'-\ell}^{(c)})^* E_c^2 .
   \label{eq:autocorr_theta}
\end{align}
Setting $\ell=0$, we obtain the variance of $\Theta_{c,m}$:
\begin{align}
    r_{\Theta}^{(c)}[0]
   = \sum_{k} (C_{0,k,k}^{(c)})^2 (Q_c - E_c^2) + \sum_{k \ne k'} |C_{0,k,k'}^{(c)}|^2 E_c^2 .
   \label{eq:var_theta}
\end{align}

\subsubsection{Second-Order Statistics for $V_{c,m}$}
Consider next the additive noise $\{V_{c,m}\}$. As $\langle X_m X_{m+\ell}\rangle=0$, the pseudo crosscorrelations $\langle V_{c,m} V_{c,m+\ell} \rangle$ are zero for all $m$ and $\ell$, so that $V_{c,m}$ is proper complex
(recall that $\langle V_{c,m} \rangle = 0$). The correlations and covariances are therefore
\begin{align}
    & r_N^{(c)}[\ell] := \langle V_{c,m} V_{c,m+\ell}^* \rangle \nonumber \\
   & = \sum_{\substack{n\ne0 \\ n\ne\ell}} \left[ \sum_{k} C_{n,k,k}^{(c)} (C_{n-\ell,k-\ell,k-\ell}^{(c)})^* 
          \, E \, Q_c \right.\nonumber \\
   & \qquad + \sum_{k,\kt \ne k} C_{n,k,k}^{(c)} (C_{n-\ell,\kt-\ell,\kt-\ell}^{(c)})^* \, E \, E_c^2 \nonumber \\
   & \qquad \left. + \sum_{k, k' \ne k} C_{n,k,k'}^{(c)} (C_{n-\ell,k-\ell,k'-\ell}^{(c)})^*  \, E \, E_c^2 \right].
   \label{eq:r_nu}
\end{align}

On the other hand, if we condition on $\{X_\ell\}=\{x_\ell\}$ then we have the time-varying covariances
\begin{align}
    & \tilde{r}_N^{(c)}[m,\ell] := \langle V_{c,m} V_{c,m+\ell}^* | \{X_\ell\}=\{x_\ell\} \rangle \nonumber \\
    & \qquad \qquad  - \langle V_{c,m} | \{X_\ell\}=\{x_\ell\} \rangle \langle V_{c,m+\ell}^* | \{X_\ell\}=\{x_\ell\} \rangle \nonumber \\
   & = \sum_{\substack{n\ne0 \\ \tilde{n}\ne0}} \left[ \sum_{k} C_{n,k,k}^{(c)} (C_{\tilde{n},k-\ell,k-\ell}^{(c)})^* 
          (Q_c - E_c^2) \right.\nonumber \\
   & \quad \left. + \sum_{k, k' \ne k} C_{n,k,k'}^{(c)} (C_{\tilde{n},k-\ell,k'-\ell}^{(c)})^*  \, E_c^2 \right]
       x_{n+m} x_{\tilde{n}+m+\ell}^* .
   \label{eq:N-cond-corr}
\end{align}
The most important terms are those with $\tilde{n}=n-\ell$.

For example, for $\ell=0$ we have
\begin{align}
    \tilde{r}_N^{(c)}[m,0]
    & = \sum_{\substack{n\ne0 \\ \tilde{n}\ne0}} \left[ \sum_{k} C_{n,k,k}^{(c)} (C_{\tilde{n},k,k}^{(c)})^* 
          (Q_c - E_c^2) \right.\nonumber \\
   & \quad \left. + \sum_{k, k' \ne k} C_{n,k,k'}^{(c)} (C_{\tilde{n},k,k'}^{(c)})^*  \, E_c^2 \right]
       x_{n+m} x_{\tilde{n}+m}^*  .
\end{align}
In other words, the variance of the noise $V_m$ depends on the previous and past symbols.

We repeat the above for the pseudo-covariances. We have the time-varying function
\begin{align}
    & \tilde{\tilde{r}}_N^{(c)}[m,\ell] := \langle V_{c,m} V_{c,m+\ell} | \{X_\ell\}=\{x_\ell\} \rangle \nonumber \\
    & \qquad \qquad  - \langle V_{c,m} | \{X_\ell\}=\{x_\ell\} \rangle \langle V_{c,m+\ell} | \{X_\ell\}=\{x_\ell\} \rangle \nonumber \\
   & = \sum_{\substack{n\ne0 \\ \tilde{n}\ne0}} \left[ \sum_{k} (-1) C_{n,k,k}^{(c)} C_{\tilde{n},k-\ell,k-\ell}^{(c)} 
          (Q_c - E_c^2) \right.\nonumber \\
   & \quad \left. - \sum_{k, k' \ne k} C_{n,k,k'}^{(c)} C_{\tilde{n},k'-\ell,k-\ell}^{(c)}  \, E_c^2 \right]
       x_{n+m} x_{\tilde{n}+m+\ell} .
   \label{eq:N-cond-pseudocorr}
\end{align}
The pseudo-covariance is not necessarily zero, and therefore the additive noise is not necessarily proper complex when conditioned on the symbol sequence $\{X_\ell\}=\{x_\ell\}$ even after subtracting off the means.

The most important terms are again those with $\tilde{n}=n-\ell$. For example, for $\ell=0$ we have
\begin{align}
    \tilde{\tilde{r}}_N^{(c)}[m,0]
    & = \sum_{\substack{n\ne0 \\ \tilde{n}\ne0}} \left[ \sum_{k} (-1) C_{n,k,k}^{(c)} C_{\tilde{n},k,k}^{(c)}
          (Q_c - E_c^2) \right.\nonumber \\
   & \quad \left. - \sum_{k, k' \ne k} C_{n,k,k'}^{(c)} C_{\tilde{n},k',k}^{(c)}  \, E_c^2 \right]
       x_{n+m} x_{\tilde{n}+m} .
\end{align}

\subsubsection{Intra-Channel Crosscorrelations}
The intra-channel crosscorrelations and pseudo-crosscorrelations are
\begin{align}
   & \langle X_m \Theta_{c,\ell} \rangle = 0 \text{ for all } \ell,m \\
   & \langle V_{c,\ell} \Theta_{c,m} \rangle = 0  \text{ for all } \ell,m \\
   & \langle X_m V_{c,m}^* \rangle = \langle X_m V_{c,m} \rangle = 0 \text{ for all } m \\
   & \langle X_m V_{c,\ell} \rangle = 0 \text{ for } \ell \ne m
\end{align}
but for $\ell \ne m$ we also have
\begin{align}
   \langle X_m V_{c,\ell}^* \rangle
   & = - j \sum_{k,k'} (C_{m-\ell,k,k'}^{(c)})^* \cdot \langle |X_m|^2 \rangle \langle B_{c,k+\ell}^*\, B_{c,k'+\ell} \rangle \nonumber \\
   & = - j \sum_{k} (C_{m-\ell,k,k}^{(c)})^* \cdot E E_c .
\end{align}
Thus, the additive noise is correlated with the $X_m$.

\subsubsection{Inter-Channel Crosscorrelations}
Consider two channels $c$ and $c'$ where $c \ne c'$. The phase noise crosscorrelations are
\begin{align}
   & \langle \Theta_{c,m} \Theta_{c',m+\ell} \rangle \nonumber \\
   & = \sum_{k,\kt} C_{0,k,k}^{(c)} C_{0,\kt,\kt}^{(c')} E_c^2
      =  \langle \Theta_{c,m} \rangle \langle \Theta_{c',m+\ell} \rangle .
\end{align}
Thus, the inter-channel phase noise processes have zero covariance and we have
\begin{equation}
    r_{\Theta}[\ell] = \sum_{c\ne0} r_{\Theta}^{(c)}[\ell] .
    \label{eq:r_Theta}
\end{equation}

We similarly compute
\begin{align}
   \langle V_{c,m} V_{c',m+\ell}^* \rangle
      = \sum_{\substack{n\ne0 \\ n\ne\ell}} \left[
      \sum_{k,\kt} C_{n,k,k}^{(c)} (C_{n-\ell,\kt-\ell,\kt-\ell}^{(c')})^* \, E \, E_c E_{c'} \right] \nonumber \\
    = \sum_{\substack{n\ne0 \\ n\ne\ell}}
      \left( \sum_{k} C_{n,k,k}^{(c)} \right) \left( \sum_{\kt} C_{n-\ell,\kt-\ell,\kt-\ell}^{(c')} \right)^* E\, E_c E_{c'}
      \label{eq:r_nu_inter}
\end{align}
and $\langle V_{c,m} V_{c',m+\ell} \rangle = 0$. In other words, the inter-channel additive noise processes are correlated and we have
\begin{equation}
    r_N[\ell] = \sum_{\substack{c\ne 0 \\ c'\ne 0}} \langle V_{c,m} V_{c',m+\ell}^* \rangle .
    \label{eq:r_N}
\end{equation}

Finally, we compute
\begin{align}
   \langle \Theta_{c,m} V_{c',m+\ell}^* \rangle = \langle \Theta_{c,m} V_{c',m+\ell} \rangle = 0
\end{align}
so the phase and additive noise processes are uncorrelated.

\subsection{Large Accumulated Dispersion}
\begin{figure}[tbp]\centering
	\setlength{\figurewidth}{0.85\linewidth}
	\setlength{\figureheight}{0.5\figurewidth}
	\input{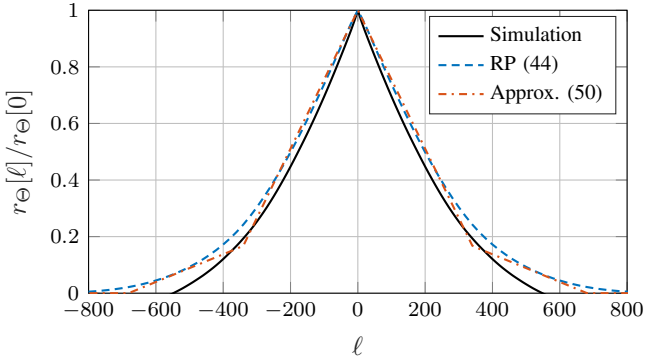}
	\caption{Normalized covariance functions of the phase noise $\{\Theta_m\}$ for a $1000$-km link with $\mathcal{P}=-6$ dBm per channel and the parameters in Table~\ref{tab:parameters} ($r_\Theta[0]=0.0034$).}
	\label{fig:autocorr}
\end{figure}

To illustrate the above derivations, consider the paper~\cite{dar2013properties} that studies links with large accumulated dispersion, i.e., $\mu_c=\left\lfloor \beta_2 \Omega_c \mathcal{L}/T\right\rfloor\gg 1$ for all $c\in\set{C}$. This paper uses the seemingly coarse approximation
\begin{equation}
    C_{0,k,k}^{(c)} \approx \begin{cases}\frac{2\gamma}{\left|\beta_2\Omega_c\right|}, & \textrm{if } 0\le -k\sign(\beta_2\Omega_c) \le \frac{\left|\beta_2\Omega_c\right|\mathcal{L}}{T}; \\
    0, & \textrm{otherwise}\end{cases}
\end{equation}
and all other $C_{0,k,k'}^{(c)}$ are approximated as zero. For $C-1$ interfering channels, we thus have (cf.~\cite{dar2013properties})
\begin{align}
  \left\langle\Theta_{m}\right\rangle & \approx 2(C-1)\gamma E_c \frac{\mathcal{L}}{T} \\
  r_{\Theta}[\ell] & \approx \frac{4\gamma^2 \mathcal{L}}{T}%
  \sum_{c\ne 0} \frac{Q_c-E_c^2}{\left|\beta_2\Omega_c\right|}\left[1-\frac{|\ell|T}{\left|\beta_2\Omega_c\right|\mathcal{L}}\right]^+. \label{eq:autocorr_theta_analytic}
\end{align}
Fig.~\ref{fig:autocorr} shows the numerical covariance $r_\Theta[\ell]$ computed from simulated data using particle filtering (see Section~\ref{sec:particle_filtering}). It can be seen that the covariance predicted by RP~\eqref{eq:r_Theta} and the approximation~\eqref{eq:autocorr_theta_analytic} are very close to the numerical covariance for a $1000$-km link with IDA and the parameters in Table~\ref{tab:parameters}.

\begin{table}[tbp]\centering
	\caption{System parameters}
	\label{tab:parameters}
	\begin{tabular}{ccc}
		\hline
		\textbf{Parameter} & \textbf{Symbol} & \textbf{Value} \\
		\hline
		Attenuation coefficient & $\alpha$ & $0.2\;\textrm{dB}/\mathrm{km}$ \\
		Dispersion coefficient & $\beta_2$ & $-21.7\;\mathrm{ps}^2/\mathrm{km}$ \\
		Nonlinear coefficient & $\gamma$ & $1.27\;\mathrm{W}^{-1}\mathrm{km}^{-1}$  \\
		Phonon occupancy factor & $\eta$ & $1$  \\
		Transmit pulse shape & $s(t)$ & sinc \\
		Number of WDM channels & $C$ & $5$ \\
		Channel bandwidth & $\mathcal{B}_{\textrm{ch}}$ & $50\;\textrm{GHz}$ \\
		Subcarrier bandwidth & $\mathcal{B}_{\textrm{sc}}$ & $8.333\;\textrm{GHz}$ \\
		Channel spacing & $\mathcal{B}_{\textrm{sp}}$ & $50\;\textrm{GHz}$ \\
		Channel of interest & $c=0$ & Center channel \\
		\hline
	\end{tabular}
\end{table}

\section{Simplified Models for Computation}
\subsection{Wiener Phase Noise (WPN) Model}
The covariance function of the phase noise $\{\Theta_m\}$ decays slowly, see Fig.~\ref{fig:autocorr}. A popular simplified model is the \emph{Wiener phase noise (WPN)} model~\cite{secondini2017fiber} where $\tilde{\Theta}_m:=\Theta_m-\langle\Theta_m\rangle$ is a discrete-time Wiener process with realizations
\begin{equation}
\tilde{\theta}_m=\tilde{\theta}_{m-1}+\sigma_\Delta\delta_m
\label{eq:wiener_update}
\end{equation}
and where the $\{\delta_m\}$ are i.i.d. real Gaussian variables with zero mean and unit variance. This model has memory $\mu=1$.

We remark that WPN seems simple and general, but it has two issues. First, WPN has only one parameter $\sigma_\Delta^2$ which does not permit to control the phase noise variance and correlation length simultaneously. Next, WPN is a non-stationary process in Euclidean (non-modulo) space, e.g., its variance grows with time. As a result, in phase (modulo) space the WPN steady-state distribution has a uniform phase irrespective of the starting phase, which does not agree with the variance~\eqref{eq:var_theta} predicted by the LP or  CPAN models.

\subsection{Markov Phase Noise (MPN) Model}
The above motivates modeling the phase noise $\{\Theta_m\}$ as a Markov chain with memory $\mu$ and conditional density:
\begin{equation}
p\left(\tilde{\theta}_m|\boldsymbol{\tilde{\theta}}_1^{m-1}\right)
= p\left(\tilde{\theta}_m|\boldsymbol{\tilde{\theta}}_{m-\mu}^{m-1}\right)
\label{eq:markov_p}
\end{equation}
where $\boldsymbol{\tilde{\theta}}_m^n$ is the vector $(\tilde{\theta}_m,\ldots,\tilde{\theta}_n)$. For each $m$, we model $(\tilde{\Theta}_{m-\mu},\ldots,\tilde{\Theta}_m)$ as jointly Gaussian with zero mean %
and with a symmetric Toeplitz covariance matrix $\mathbf{C}_\mu$ whose first column is $(r_\Theta[0], \ldots, r_\Theta[\mu])^T$ from~\eqref{eq:r_Theta} where the superscript $T$ denotes transposition.

The conditional distribution~\eqref{eq:markov_p} can be computed from the mean vector and covariance matrix~\cite[Ch. 2, Sec. 3.4]{eaton1983multivariate}. The result is that $\tilde{\Theta}_m|\tilde{\boldsymbol{\theta}}_{1}^{m-1}$ is Gaussian with mean
\begin{align}
    \left\langle \tilde{\Theta}_m \left| \tilde{\boldsymbol{\theta}}_{1}^{m-1} \right. \right\rangle = 
    \mathbf{g}_\mu \left(\tilde{\boldsymbol{\theta}}_{m-\mu}^{m-1}\right)^T
\end{align} and variance $\sigma_{\mu}^2$, where
\begin{subequations}
	\begin{align}
	& \mathbf{g}_\mu=\left(r_\Theta[\mu],\ldots,r_\Theta[1]\right)\left(\mathbf{C}_{\mu-1}\right)^{-1} \\
	& \sigma_{\mu}^2=r_\Theta[0]-\mathbf{g}_\mu\left(r_\Theta[\mu],\ldots,r_\Theta[1]\right)^T.
	\end{align}
	\label{eq:markov_parameters}
\end{subequations}
This yields the recursive \emph{Markov phase noise} (MPN) model
\begin{equation}
\tilde{\theta}_m = \mathbf{g}_\mu \left(\tilde{\boldsymbol{\theta}}_{m-\mu}^{m-1}\right)^T + \sigma_{\mu} \delta_m
\label{eq:markov_update}
\end{equation}
where the $\{\Delta_m\}$ are i.i.d. real Gaussian variables with zero mean and unit variance. Note that $\sigma_{\mu}^2$ does not depend on $\{\theta_m\}$. Note also that we perform the computations \eqref{eq:markov_parameters}-\eqref{eq:markov_update} in Euclidean (non-modulo) space.

For example, for memory $\mu=1$, the MPN model has
\begin{equation}
\tilde{\theta}_m=\frac{r_\Theta[1]}{r_\Theta[0]}\tilde{\theta}_{m-1}+\sqrt{r_\Theta[0]-\frac{r_\Theta[1]^2}{r_\Theta[0]}}\delta_m
\end{equation}
which is different than~\eqref{eq:wiener_update}, e.g., $\langle \tilde{\Theta}_m^2 \rangle$ does not increase with $m$. %

\subsection{CPAN Model with Simplified Memory}
\begin{figure}[tbp]\centering
	\setlength{\figurewidth}{0.85\linewidth}
	\setlength{\figureheight}{0.4\figurewidth}
	\definecolor{mycolor1}{rgb}{0.00000,0.44700,0.74100}%
\begin{tikzpicture}

\begin{axis}[%
width=0.951\figurewidth,
height=\figureheight,
at={(0\figurewidth,0\figureheight)},
scale only axis,
xmin=-10,
xmax=10,
xlabel style={font=\color{white!15!black}},
xlabel={$\ell$},
ylabel={$r_Z[\ell]\ (\textrm{Watt}\cdot\textrm{s})$},
ymin=-4e-18,
ymax=1.2e-17,
grid,
axis background/.style={fill=white},
legend style={legend cell align=left, align=left, draw=white!15!black}
]
\addplot [color=mycolor1, thick, forget plot]
  table[row sep=crcr]{%
-10	1.76341770901977e-19\\
-9	-2.15525417453547e-19\\
-8	2.5537724002031e-19\\
-7	-2.80692194240514e-19\\
-6	3.62197481222776e-19\\
-5	-4.41287150609291e-19\\
-4	5.86652062332491e-19\\
-3	-8.03762511075025e-19\\
-2	1.17721167279488e-18\\
-1	-2.1874119051756e-18\\
0	1.03104670112047e-17\\
1	-2.1874119051756e-18\\
2	1.17721167279488e-18\\
3	-8.03762511075025e-19\\
4	5.86652062332491e-19\\
5	-4.41287150609291e-19\\
6	3.62197481222776e-19\\
7	-2.80692194240514e-19\\
8	2.5537724002031e-19\\
9	-2.15525417453547e-19\\
10	1.76341770901977e-19\\
};

\end{axis}

\end{tikzpicture}%
	\caption{Real part of the autocorrelation function of the residual additive noise $w_m+\nu_m$ in the center channel of a $1000$-km link with the parameters in Table~\ref{tab:parameters}. Channel power: $\mathcal{P}=-6$ dBm. Symbol energy: $E=\mathcal{P}T=5.02\cdot{10}^{-15} \textrm{W}\cdot\textrm{s}.$}
	\label{fig:autocorr_w}
\end{figure}
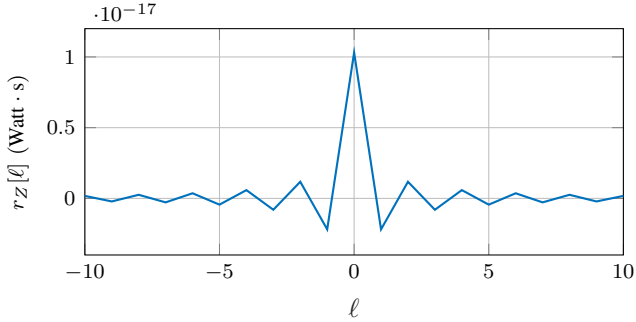

We combine the ASE noise $w_m$ and the NLI noise $v_m$ in one correlated noise term $z_m$. The simplified CPAN model is
\begin{equation}
y_m=x_m e^{j\theta_m}+z_m
\label{eq:cpan_model2}
\end{equation}
where $\theta_m=\tilde{\theta}_m+\langle \Theta_m \rangle$ and $\tilde{\theta}_m$ is a real zero-mean Gaussian Markov process generated according to~\eqref{eq:markov_update}, and $z_m$ is circularly-symmetric Gaussian with autocorrelation function
\begin{equation}
r_Z[\ell]\triangleq\left\langle Z_m Z_{m+\ell}^* \right\rangle=\sigma_W^2\delta[\ell]+ r_N[\ell]
\label{eq:autocorr_w}
\end{equation}
where $\sigma_W^2=N_{\textrm{ASE}}$, and $r_N[\ell]$ is given by~\eqref{eq:r_N}. Fig.~\ref{fig:autocorr_w} shows the simulated autocorrelation function of $\{Z_m\}$ in the center channel of a $5$-channel WDM system with $\mathcal{L}=1000$ km and the parameters in Table~\ref{tab:parameters}. With these parameters, the imaginary part of the autocorrelation function turns out to be negligible.

\section{Achievable Rates and Particle Filters}
Given an input distribution $p(x)$, an achievable rate (lower bound on capacity) of a channel with conditional distribution $p(y|x)$ can be obtained via a \textit{mismatched model} $q(y|x)$ as
\begin{equation}
I_q(X;Y)=\left\langle \log_2 \frac{q(Y|X)}{q(Y)}\right\rangle\le I(X; Y)
\label{eq:mismatched-lower-bound}
\end{equation}
where $q(y)=\int p(x)q(y|x)\diff{x}$ is the mismatched output distribution, and where the expectation is performed with respect to the true distribution $p(x,y)$. The expectation can therefore be computed by simulations. One usually chooses $q(y|x)$ to approach $p(y|x)$ while at the same time being simple enough to compute.

To develop an auxiliary channel model, we first compensate for $\langle{\Theta_m}\rangle$ and whiten the noise $\{Z_m\}$. We use a real and normalized whitening filter $\mathbf{h}=(h_{0},\dots,h_{L-1})$ with $L$ taps and $\|\mathbf{h}\|^2=1$ and compute
\begin{align}
u_m &= e^{-j\langle{\Theta_m}\rangle} \sum_{\ell=0}^{L-1} h_\ell \, y_{m-\ell} \nonumber \\
   & = \sum_{\ell=0}^{L-1} h_\ell \, x_{m-\ell} \, e^{j\tilde{\theta}_{m-
   \ell}} + \tilde{z}_m.
\label{eq:u_n}
\end{align}
We consider the channel from the $\{X_m\}$ to the $\{U_m\}$, i.e., our mismatched model is given by~\eqref{eq:u_n} assuming that the $\{\tilde{Z}_m\}$ are i.i.d. Gaussian with variance $\sigma_{Z}^2$. In the rate computations, this variance is estimated from simulated data using a maximum-likelihood estimator, see~\eqref{eq:var_w_estimator} below.

\subsection{Mismatched Output Density $q(\mathbf{u})$}
The input symbols $\{X_m\}$ are circularly symmetric i.i.d. Gaussian with variance $E=\mathcal{P}T$. The variables $\{X_m e^{j\tilde{\Theta}_m}\}$ are thus also circularly symmetric i.i.d. Gaussian with variance $E=\mathcal{P}T$. This in turn implies that the $\{U_m\}$ in~\eqref{eq:u_n} are Gaussian with zero mean and autocorrelation function
\begin{equation}
    r_U[\ell] = \left\langle U_m U_{m+\ell}^*\right\rangle = \sum_{k=0}^{L-1} h_k h_{k+\ell}^* E + \sigma_Z^2\,\delta[\ell].
\end{equation}
Consider the vector $\mathbf{u}=(u_1,\ldots,u_M)^T$ of $M$ output symbols with density
\begin{align}
    q\left(\mathbf{u}\right) = \frac{1}{\det\left(\pi\mathbf{R}_U\right)} \exp\left( -\mathbf{u}^H \mathbf{R}_U^{-1}\mathbf{u} \right)
    \label{eq:q_U}
\end{align}
where the covariance matrix $\mathbf{R}_U$ is an $M\times M$ Toeplitz matrix whose first row is $(r_U[0],\ldots,r_U[M-1])$ and whose first column is $(r_U[0],\ldots,r_U[-M+1])$, and where $\det(\pi\mathbf{R}_U)$ is the determinant of $\pi \mathbf{R}_U$. For small $L$, this is a sparse band matrix and~\eqref{eq:q_U} can be computed efficiently. Our upper bound on the output entropy is
\begin{align}
    h_q(U)
    =\left\langle -\log_2 q\left(\mathbf{U}\right) \right\rangle
    \approx \frac{1}{N} \sum_{n=1}^N -\log_2 q(\mathbf{u}_n)
\end{align}
where $\{\mathbf{u}_n\}$ is generated by $N$ independent simulation runs.

\subsection{Mismatched Channel Density $q(\mathbf{u}|\mathbf{x})$}\label{sec:particle_filtering}

Similar to~\cite{secondini2017fiber, secondini2019nonlinearity}, we apply particle filtering~\cite{dauwels2008particle} to~\eqref{eq:u_n} to estimate
\begin{align}
     h_q(U|X)=\left\langle - \log_2 q(\mathbf{U}|\mathbf{X})\right\rangle.
\end{align}

A particle filter tracks the phase noise \emph{state} vector $\boldsymbol{\tilde{\theta}}_{m-\mu}^{m-1}$ with a list of $K$ particles that is updated for every new received symbol $u_m$. More precisely, after processing $u_{m-1}$, the $k$-th \textit{particle}, $k=1,\dots,K$, is an ordered pair of a realization
$\left(\hat{\theta}_{m-\mu}^{(k)},\ldots,\hat{\theta}_{m-1}^{(k)}\right)$ of $\boldsymbol{\tilde{\theta}}_{m-\mu}^{m-1}$
and an \emph{importance weight} $W_{m-1}^{(k)}$ that approximates the relative posterior density $p(\boldsymbol{\tilde{\theta}}_{m-\mu}^{m-1}|\mathbf{x},\mathbf{u}_{1}^{m-1})$ such that the weights sum to one:
\begin{align}
    \sum_{k=1}^K W_{m-1}^{(k)} = 1.
\end{align}

Upon receiving $u_m$, the particle list is updated by performing the following steps.
\begin{enumerate}
\item Update the $K$ realizations: for each $k$, $k=1,\ldots,K$, generate $\hat{\theta}_m^{(k)}$ via~\eqref{eq:markov_update} (or~\eqref{eq:wiener_update} for the WPN model). The goal is that $\hat{\theta}_m^{(k)}$ is distributed according to~\eqref{eq:markov_p}:
\begin{equation}
	\hat{\theta}_m^{(k)} \sim p_{\tilde{\Theta}_m|\boldsymbol{\tilde{\Theta}}_{m-\mu}^{m-1}} \left( \cdot \left| \: \hat{\theta}_{m-\mu}^{(k)},\dots,
	\hat{\theta}_{m-1}^{(k)} \right. \right).
\end{equation}
	
\item Estimate $q(u_m | \mathbf{x},\mathbf{u}_1^{m-1})$:
We have
\begin{align}
    & q(u_m | \mathbf{x},\mathbf{u}_1^{m-1}) =
    \int p\left( \left. u_m, \boldsymbol{\tilde{\theta}}_{m-\mu}^{m} \right| \mathbf{x},\mathbf{u}_1^{m-1} \right) \, d\boldsymbol{\tilde{\theta}}_{m-\mu}^{m} \nonumber \\
    & \quad = \left\langle \left.
    q\left( \left. u_m \right| \mathbf{x},\mathbf{u}_1^{m-1}, \boldsymbol{\tilde{\Theta}}_{m-\mu}^{m}  \right) \right| \mathbf{x},\mathbf{u}_1^{m-1}
    \right\rangle
    \label{eq:PF-q-estimate}
\end{align}
where the expectation is with respect to
\begin{align} 
    & p\left( \left. \boldsymbol{\tilde{\theta}}_{m-\mu}^{m} \right| \mathbf{x},\mathbf{u}_1^{m-1} \right) \nonumber \\
    & \quad = p\left( \left. \boldsymbol{\tilde{\theta}}_{m-\mu}^{m-1} \right| \mathbf{x},\mathbf{u}_1^{m-1} \right) \cdot p\left( \tilde{\theta}_{m} \left| \boldsymbol{\tilde{\theta}}_{m-\mu}^{m-1} \right. \right).
    \label{eq:PF-density-factors}
\end{align}
The expressions \eqref{eq:PF-q-estimate} and \eqref{eq:PF-density-factors} motivate computing the average
\begin{equation}
	D_m=\sum_{k=1}^{K} W_{m-1}^{(k)} \underbrace{p_{\tilde{Z}}\left(u_{m} - \sum_{\ell=0}^{L-1} h_\ell x_{m-\ell} e^{j\hat{\theta}_{m-\ell}^{(k)}} \right)}_{q\left( u_m \left| \mathbf{x},\mathbf{u}_1^{m-1}, \tilde{\theta}_{m-\mu}^{(k)},\dots,
	\tilde{\theta}_{m}^{(k)} \right. \right)}
\end{equation}
where $\tilde{Z}_m$ is circularly-symmetric Gaussian with variance $\sigma_Z^2$. Note that this step requires $\mu \ge L-1$ because we need the $L-1$ past samples of the phase noise (alternatively, one can store more of the past of each particle). By an argument similar to~\cite[Eqs. (14)-(16)]{dauwels2008particle}, for large $K$ the value $D_m$ is approximately \eqref{eq:PF-q-estimate}.
	
\item Update the weights:
\begin{equation}
	W_m^{(k)}= W_{m-1}^{(k)} \frac{1}{D_m} p_{\tilde{Z}}\left(u_{m}-\sum_{\ell=0}^{L-1} h_\ell x_{m-\ell} e^{j\hat{\theta}_{m-\ell}^{(k)}}\right)
\end{equation}
for $k=1,\dots,K$. Note that $\sum_k W_m^{(k)}=1$. %
	
\item The particle list tends to degenerate, i.e., to concentrate the weight on one particle, while the weight of all other particles becomes negligible. The effective number of particles can be measured, e.g., by counting the number of particles with a probability greater than some threshold. Another heuristic is to define the effective number of particles as (see, e.g.,~\cite{dauwels2008particle})
\begin{equation}
	K_{\textrm{eff}}\triangleq\frac{1}{\sum_k \left(W_{m}^{(k)}\right)^2}.
\end{equation} 
Note that $K_{\textrm{eff}}$ takes on its largest value $K$ when $W_{m}^{(k)}=1/K$ for all $k$, and its minimum value 1 if there is one particle with positive probability. Thus, for some specified $\epsilon<1$, if $K_{\textrm{eff}}$ becomes smaller than $\epsilon K$ then we resample the particles by drawing $K$ new realizations from $\left\{\left(\theta_{m-\mu+1}^{(k)},\ldots, \theta_{m}^{(k)}\right)\right\}_{k=1}^{K}$ with probabilities $W_m^{(k)}$. Then set all $W_m^{(k)}$ to $1/K$. As suggested in~\cite{dauwels2008particle}, we use $\epsilon=0.3$.
\end{enumerate}

After the last iteration, our upper bound on the conditional entropy is
\begin{equation}
h_q(U|X) \approx \sum_{m=1}^M -\log_2 D_m.
\end{equation}
For $N$ independent simulation runs, we compute
\begin{equation}
h_q(U|X) \approx \frac{1}{N} \sum_{n=1}^N \sum_{m=1}^M -\log_2 D_m^{(n)}
\end{equation}
where $D_m^{(n)}$ is the $D_m$ of the $n$th simulation.

Finally, our lower bound on the capacity of the NLSE is
\begin{align}
I_q(X; U) & = h_q(U)-h_q(U|X) \nonumber \\
  & \overset{(a)}{\le} I(X; U)
    \overset{(b)}{\le} I(X; Y)
\label{eq:rate}
\end{align}
where step $(a)$ follows by \eqref{eq:mismatched-lower-bound} and step $(b)$ follows by the data processing inequality.

\section{Estimating Model Parameters}
We estimate the parameters of the simplified CPAN model~\eqref{eq:cpan_model2} from simulated data in a training phase.  Similar to~\cite{secondini2019nonlinearity}, we use a maximum-likelihood estimator for the additive noise variance based on $\{|y_m|\}$ and $\{|x_m|\}$:
\begin{equation}
\hat{\sigma}_Z^2=\arg\max_{\sigma^2}\sum_{m=1}^M\log L(|y_m|, |x_m|;\sigma^2)
\label{eq:var_w_estimator}
\end{equation}
where the likelihood function is a Rice density
\begin{equation}
L(|y_m|, |x_m|;\sigma^2)
=\frac{2|y_m|}{\sigma^2}e^{-\frac{|y_m|^2+|x_m|^2}{\sigma^2}}I_0\left(\frac{2|y_m||x_m|}{\sigma^2}\right).
\end{equation}
From~\eqref{eq:cpan_model2}, we have
\begin{align}
\langle Y_m X_m^* \rangle
  & =\langle |X_m|^2\rangle
  \cdot \langle e^{j \tilde{\Theta}_m} \rangle
  \cdot e^{j\langle\Theta_m\rangle}
\end{align}
and $\langle e^{j \tilde{\Theta}_m} \rangle$ is real because $\tilde{\Theta}_m$ and $-\tilde{\Theta}_m$ have the same statistics.
We can thus estimate $\langle\Theta_m\rangle$ via
\begin{align}
\langle\hat\Theta_m\rangle
= \mathrm{angle}\left(\frac{1}{M}\sum_{m=1}^{M}y_mx_m^*\right).
\label{eq:E_theta_estimator}
\end{align}
For the MPN model, we choose $\hat{r}_\Theta[0]\ldots \hat{r}_\Theta[\mu]$ by assuming that $r_\Theta[\ell]$ is a scaled version of~\eqref{eq:autocorr_theta_analytic} and by minimizing $h(Y|X)$ over the scaling factor, where $h(Y|X)$ is computed using the particle method. We use these estimated $\hat{r}_\Theta[0]\ldots \hat{r}_\Theta[\mu]$ in~\eqref{eq:markov_update} to compute the parameters $\mathbf{g}_\mu$ and $\sigma_v$ of the MPN model.

We use a real symmetric whitening filter with $L=3$ taps:
\begin{equation}
\mathbf{h}=\left(h_2, \quad\sqrt{1-2h_2^2},\quad h_2\right).
\label{eq:h}
\end{equation}
We estimate $h_2$ by minimizing the cost function $h_q(U|X)$, which is computed using~\eqref{eq:u_n} and particle filtering. Increasing the number of taps above $3$ did not improve the rate bounds in our simulations.

In a subsequent testing phase, the achievable rate is computed using~\eqref{eq:rate} with the estimated model parameters $\hat{\sigma}_Z^2$, $\hat{r}_{\Theta}[\ell]$, and $\langle\hat{\Theta}_m\rangle$ on a new set of simulated data.

\section{Multi-carrier Communication}

\begin{figure}
    \centering
    	\setlength{\figurewidth}{0.4\columnwidth}
	\setlength{\figureheight}{1.0\figurewidth}
    \definecolor{mycolor1}{rgb}{0.00000,0.44700,0.74100}%
\definecolor{mycolor2}{rgb}{0.85000,0.32500,0.09800}%
\definecolor{mycolor3}{rgb}{0.92900,0.69400,0.12500}%
\definecolor{mycolor4}{rgb}{0.49400,0.18400,0.55600}%
\definecolor{mycolor5}{rgb}{0.46600,0.67400,0.18800}%
\definecolor{mycolor6}{rgb}{0.30100,0.74500,0.93300}%
\definecolor{mycolor7}{rgb}{0.63500,0.07800,0.18400}%
\begin{tikzpicture}

\begin{axis}[%
width=0.951\figurewidth,
height=\figureheight,
at={(0\figurewidth,0\figureheight)},
scale only axis,
xmin=-10,
xmax=-4,
xlabel={$S\cdot\mathcal{P}_s$, in dBm},
xtick distance=1,
ymin=0,
ymax=0.014,
ytick distance=2e-3,
grid,
axis background/.style={fill=white},
title={(a) $r_{\Theta}[0]$},
legend style={at={(0.00, 1)}, anchor=north west, legend cell align=left, align=left, draw=white!15!black, inner sep=0pt}
]
\addplot [thick, color=black]
  table[row sep=crcr]{%
-13	0.000185740536864854\\
-12	0.000309137747275814\\
-11	0.000466696740282207\\
-10	0.0007083821310532\\
-9	0.00105526263450439\\
-8	0.00187427748519892\\
-7	0.00311571586030556\\
-6	0.00482452639582846\\
-5	0.00785540279836134\\
-4	0.0133645265526386\\
};
\addlegendentry{Single-carrier}

\addplot [thick, color=mycolor1, mark=triangle]
  table[row sep=crcr]{%
-13	0.000185593371354101\\
-12	0.000305183768941004\\
-11	0.000415355924525169\\
-10	0.000806472003972431\\
-9	0.00109953585037448\\
-8	0.00160461861867571\\
-7	0.00276336560160283\\
-6	0.00368940474044069\\
-5	0.00591126690352994\\
-4	0.0102170803231851\\
};
\addlegendentry{Subcarrier  1}

\addplot [thick, color=mycolor2, mark=square]
  table[row sep=crcr]{%
-13	0.000166203272405562\\
-12	0.000332062294361344\\
-11	0.000383625555377173\\
-10	0.000505932010652835\\
-9	0.000703790372458355\\
-8	0.00105635548723798\\
-7	0.00162486684554479\\
-6	0.00205974905136203\\
-5	0.00271247985831934\\
-4	0.00424630665969398\\
};
\addlegendentry{Subcarrier  2}

\addplot [thick, color=mycolor3, mark=o]
  table[row sep=crcr]{%
-13	0.000187521089286749\\
-12	0.000212853787098241\\
-11	0.00033736921056255\\
-10	0.00051913040877283\\
-9	0.000618275494551628\\
-8	0.000789722682964613\\
-7	0.00122183681406653\\
-6	0.00168852466877981\\
-5	0.00236058467209017\\
-4	0.00343343698438505\\
};
\addlegendentry{Subcarrier  3}

\end{axis}
\end{tikzpicture}%
    \definecolor{mycolor1}{rgb}{0.00000,0.44700,0.74100}%
\definecolor{mycolor2}{rgb}{0.85000,0.32500,0.09800}%
\definecolor{mycolor3}{rgb}{0.92900,0.69400,0.12500}%
\definecolor{mycolor4}{rgb}{0.49400,0.18400,0.55600}%
\definecolor{mycolor5}{rgb}{0.46600,0.67400,0.18800}%
\definecolor{mycolor6}{rgb}{0.30100,0.74500,0.93300}%
\definecolor{mycolor7}{rgb}{0.63500,0.07800,0.18400}%
\begin{tikzpicture}

\begin{axis}[%
width=0.951\figurewidth,
height=\figureheight,
at={(0\figurewidth,0\figureheight)},
scale only axis,
xmin=-10,
xmax=-4,
xtick distance=1,
ymin=8,
ymax=9.5,
grid,
ytick distance={0.2},
xlabel={$S\cdot\mathcal{P}_s$, in dBm},
title={(b) SE (bits/s/Hz)},
axis background/.style={fill=white},
legend style={legend cell align=left, align=left, draw=white!15!black}
]
\addplot [thick, color=black]
  table[row sep=crcr]{%
-13	7.3932257\\
-12	7.7098327\\
-11	8.0171282\\
-10	8.2979785\\
-9	8.5469007\\
-8	8.7339973\\
-7	8.8270297\\
-6	8.8097448\\
-5	8.6196400\\
-4	8.2896925\\
};
\addlegendentry{Single-carrier}

\addplot [thick, color=mycolor1, mark=triangle]
  table[row sep=crcr]{%
-13	7.33285369072277\\
-12	7.64254555212907\\
-11	7.92037955596366\\
-10	8.18087509706381\\
-9	8.3793068035992\\
-8	8.54814876858911\\
-7	8.64159787405378\\
-6	8.60748496365524\\
-5	8.45317891393584\\
-4	8.14853116945705\\
};
\addlegendentry{data1}

\addplot [thick, color=mycolor2, mark=square]
  table[row sep=crcr]{%
-13	7.37517855462278\\
-12	7.69747344090066\\
-11	7.99974587907369\\
-10	8.29805078692491\\
-9	8.56754095113421\\
-8	8.82065656314815\\
-7	9.02053937827208\\
-6	9.15690383278068\\
-5	9.20294930420968\\
-4	9.13756560838158\\
};
\addlegendentry{data2}

\addplot [thick, color=mycolor3, mark=o]
  table[row sep=crcr]{%
-13	7.38111429893145\\
-12	7.70032969512048\\
-11	8.01364128552803\\
-10	8.30058296670247\\
-9	8.59007424149744\\
-8	8.85229574418113\\
-7	9.07815581939611\\
-6	9.24764623930722\\
-5	9.34210393577725\\
-4	9.3188832433293\\
};
\addlegendentry{data3}

\legend{};

\end{axis}

\end{tikzpicture}%
    \caption{(a) Phase-noise variance in a $1000$-km link with 6 subcarriers, uniform power allocation, and the parameters in Table~\ref{tab:parameters}. (b) Spectral efficiency of each subcarrier using particle filtering and the CPAN model. The horizontal axis is the subcarrier power times the number of subcarriers. Due to symmetry, the variances and rates for subcarriers 4, 5, and 6 are very close to those for subcarriers 3, 2, and 1 respectively. The legend is valid for both graphs.}
    \label{fig:pn_variance_sc}
\end{figure}
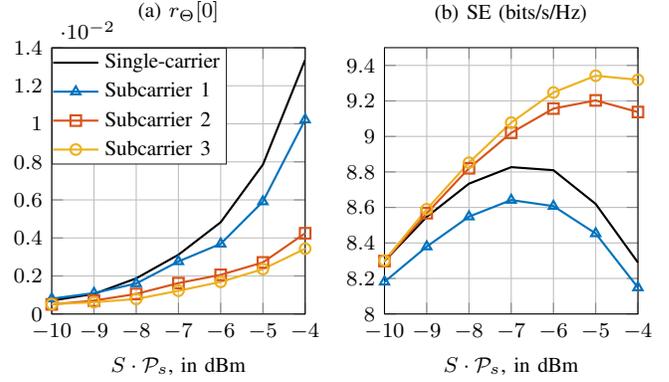

The phase noise statistics of the CPAN model are frequency-dependent. For example, consider the digitally back-propagated center channel ($c=0$) with bandwidth $\mathcal{B}_{\textrm{ch}}$. We approximate the auto-correlation function of the phase noise at angular frequency $\Omega$ by~\eqref{eq:autocorr_theta_analytic} and replace $\Omega_c$ with $\Omega_c-\Omega$, where $\Omega$ is limited to $|\Omega|\le\pi\mathcal{B}_{\textrm{ch}}$. Due to the terms $|\beta_2(\Omega_c-\Omega)|$ in the denominator, the variance of the phase noise is largest at the edges of the center channel.

To treat such frequency dependencies, the authors of~\cite{secondini2017fiber} use a multi-carrier system where each channel has $S$ subcarriers of bandwidth $\mathcal{B}_{\textrm{sc}}=\mathcal{B}_{\textrm{ch}}/S$. These $S$ subcarriers are digitally back-propagated jointly and then each subcarrier is processed independently by a particle filter. In Fig.~\ref{fig:pn_variance_sc} (a), we plot the phase noise variance $r_{\Theta}[0]$ for each subcarrier of a simulated 6-subcarrier system ($S=6$). Observe that $r_{\Theta}[0]$ is largest for subcarriers 1 and 6 at the channel band edges.

The paper~\cite{secondini2017fiber} showed that using 6 subcarriers improves the capacity by 0.2 bits/s/Hz as compared to using a single-carrier. We adopt this approach and apply frequency-dependent power allocation (FDPA) to give more power to the center subcarriers that have better channels. To choose the powers, we simulated a 6-subcarrier (6SC) system with uniform power allocation and computed the achievable rate of each subcarrier as a function of power, obtaining the curves in Fig.~\ref{fig:pn_variance_sc} (b). For the FDPA system, we used these curves as utility functions to maximize the rate:
\begin{equation}
    \max_{\{\mathcal{P}_1,\dots,\mathcal{P}_S\}} \: \sum_{s=1}^S \textrm{rate}_s(\mathcal{P}_s) \quad \textrm{s.t. } \sum_{s=1}^S \mathcal{P}_s=\mathcal{P}
    \label{eq:fdpa}
\end{equation}
where $\mathcal{P}_s$ is the power level of subcarrier $s$ for $s=1,\dots,S$,  and
where $\textrm{rate}_s(\cdot)$ is the rate function of the $s$-th subcarrier given by the corresponding curve in Fig.~\ref{fig:pn_variance_sc}(b).

FDPA is related to classic water-filling for linear channels. However, water-filling is not necessarily optimal because the subcarrier rate functions $\textrm{rate}_s(\cdot)$, $s=1,\dots,S$, are different than for the linear case. A heuristic approach to optimize the power levels is, e.g., to iteratively perform the optimization in~\eqref{eq:fdpa} by applying the rate functions obtained from the previous iteration. However, in our simulations the only noticeable rate increase was due to the first iteration. 

\section{Numerical results}
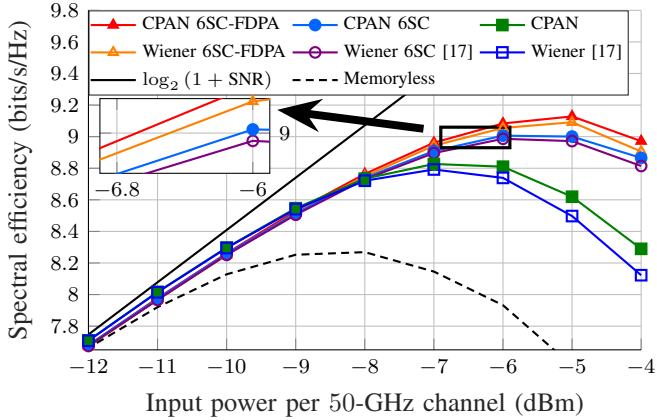
\begin{figure}[tbp]\centering
	\setlength{\figurewidth}{0.85\columnwidth}
	\setlength{\figureheight}{0.6\figurewidth}
	\definecolor{mycolor1}{rgb}{0.80000,1.00000,0.00000}%
\definecolor{mycolor2}{rgb}{0, 0.5, 0}%
\definecolor{mycolor3}{rgb}{0.00000,0.40000,1.00000}%
\definecolor{mycolor4}{rgb}{0.80000,0.00000,1.00000}%
\definecolor{mycolor5}{rgb}{1.00000,0.00000,0.00000}%
\definecolor{mycolor6}{rgb}{1.00000,0.50000,0.00000}%
\begin{tikzpicture}

\begin{axis}[%
width=0.976\figurewidth,
height=\figureheight,
at={(0\figurewidth,0\figureheight)},
scale only axis,
xmin=-12,
xmax=-4,
xlabel style={font=\color{white!15!black}},
xlabel={Input power per $50$-GHz channel (dBm)},
ymin=7.65,
ymax=9.8,
ylabel style={font=\color{white!15!black}},
ylabel={Spectral efficiency (bits/s/Hz)},
axis background/.style={fill=white},
axis x line*=bottom,
axis y line*=left,
xtick distance=1,
ytick distance=0.2,
grid,
legend style={font=\scriptsize, at={(0, 1)}, anchor=north west, legend cell align=left, align=left, draw=white!15!black, inner sep=0.1mm},
legend columns=3,
transpose legend
]

\addplot [thick, color=mycolor5, mark=triangle*, mark options={solid, mycolor5}]
  table[row sep=crcr]{%
-13	7.3730830\\
-12	7.6788651\\
-11	7.9771342\\
-10	8.2613606\\
-9	8.5292504\\
-8	8.7637648\\
-7	8.9615991\\
-6	9.0834988\\
-5	9.1282490\\
-4	8.9716717\\
};
\addlegendentry{CPAN 6SC-FDPA}

\addplot [thick, color=mycolor6, mark=triangle, mark options={solid, mycolor6}]
  table[row sep=crcr]{%
-13	7.3670431\\
-12	7.6712543\\
-11	7.9674152\\
-10	8.2500324\\
-9	8.5170728\\
-8	8.7487491\\
-7	8.9426414\\
-6	9.0551128\\
-5	9.0914072\\
-4	8.9060405\\
};
\addlegendentry{Wiener 6SC-FDPA}

\addplot [thick, color=black]
table[row sep=crcr]{%
	-13	7.41649966776195\\
	-12	7.74695458877323\\
	-11	8.07776545091807\\
	-10	8.40885959646594\\
	-9	8.74017910952567\\
	-8	9.0716778580849\\
	-7	9.40331911732814\\
	-6	9.73507366456607\\
	-5	10.0669182550719\\
	-4	10.3988344044816\\
};
\addlegendentry{$\log_2\left(1+\textrm{SNR}\right)$}

\addplot [thick, color=mycolor3, mark=*, mark options={solid, mycolor3}]
  table[row sep=crcr]{%
-13	7.3669019\\
-12	7.6802883\\
-11	7.9738735\\
-10	8.2586354\\
-9	8.5141583\\
-8	8.7384928\\
-7	8.9094252\\
-6	9.0063525\\
-5	9.0006969\\
-4	8.8652663\\
};
\addlegendentry{CPAN 6SC}

\addplot [thick, color=violet, mark=o, mark options={solid, violet, rotate=180}]
table[row sep=crcr]{%
	-13	7.3617299\\
    -12	7.6742852\\
    -11	7.9658629\\
    -10	8.2512127\\
    -9	8.5052656\\
    -8	8.7296929\\
    -7	8.8959932\\
    -6	8.9864593\\
    -5	8.9718213\\
    -4	8.8142074\\
};
\addlegendentry{Wiener 6SC~\cite{secondini2017fiber}}

\addplot [thick, color=black, densely dashed, mark options={solid, black}]
  table[row sep=crcr]{%
-13	7.37242819562934\\
-12	7.66262110187144\\
-11	7.92024380254114\\
-10	8.12830390519092\\
-9	8.25221896889787\\
-8	8.26870388455875\\
-7	8.14562733969801\\
-6	7.93260566583533\\
-5	7.55039959198644\\
-4	7.06258500714054\\
};
\addlegendentry{Memoryless}

\addplot [thick, color=mycolor2, mark=square*, mark options={solid, mycolor2}]
  table[row sep=crcr]{%
-13	7.3932\\
-12	7.7098\\
-11	8.0171\\
-10	8.2980\\
-9	8.5469\\
-8	8.7340\\
-7	8.8270\\
-6	8.8097\\
-5	8.6196\\
-4	8.2897\\
};
\addlegendentry{CPAN}

\addplot [thick, color=blue, mark=square, mark options={solid, blue}]
table[row sep=crcr]{%
	-13	7.39119791767129\\
	-12	7.70762937898945\\
	-11	8.01463070787953\\
	-10	8.2948149004191\\
	-9	8.54118046603565\\
	-8	8.72077951549698\\
	-7	8.79217012070618\\
	-6	8.7388894033506\\
	-5	8.49723674779818\\
	-4	8.12270195424516\\
};
\addlegendentry{Wiener~\cite{secondini2017fiber}}

\draw [very thick] (-6.9, 8.93) rectangle (-5.9, 9.06);
\draw [line width=3pt, -{Stealth}] (-7.15, 9.05) -- (-9.27, 9.17);

\end{axis}

\begin{axis}[%
width=0.3\figurewidth,
height=0.22\figureheight,
at={(0.02\figurewidth, 0.52\figureheight)},
scale only axis,
xmin=-6.9,
xmax=-5.9,
ymin=8.93,
ymax=9.06,
axis background/.style={fill=white},
xtick={-6.8, -6},
ytick={9},
yticklabel pos=right,
grid,
]

\addplot [forget plot, thick, color=mycolor5, mark=triangle*, mark options={solid, mycolor5}]
  table[row sep=crcr]{%
-13	7.3730830\\
-12	7.6788651\\
-11	7.9771342\\
-10	8.2613606\\
-9	8.5292504\\
-8	8.7637648\\
-7	8.9615991\\
-6	9.0834988\\
-5	9.1282490\\
-4	8.9716717\\
};

\addplot [forget plot, thick, color=mycolor6, mark=triangle, mark options={solid, mycolor6}]
  table[row sep=crcr]{%
-13	7.3670431\\
-12	7.6712543\\
-11	7.9674152\\
-10	8.2500324\\
-9	8.5170728\\
-8	8.7487491\\
-7	8.9426414\\
-6	9.0551128\\
-5	9.0914072\\
-4	8.9060405\\
};

\addplot [forget plot, thick, color=violet, mark=o, mark options={solid, violet, rotate=180}]
table[row sep=crcr]{%
	-13	7.3617299\\
    -12	7.6742852\\
    -11	7.9658629\\
    -10	8.2512127\\
    -9	8.5052656\\
    -8	8.7296929\\
    -7	8.8959932\\
    -6	8.9864593\\
    -5	8.9718213\\
    -4	8.8142074\\
};

\addplot [forget plot, thick, color=mycolor3, mark=*, mark options={solid, mycolor3}]
  table[row sep=crcr]{%
-13	7.3669019\\
-12	7.6802883\\
-11	7.9738735\\
-10	8.2586354\\
-9	8.5141583\\
-8	8.7384928\\
-7	8.9094252\\
-6	9.0063525\\
-5	9.0006969\\
-4	8.8652663\\
};

\end{axis}

\end{tikzpicture}%
	\caption{Achievable rates for a $1000$-km IDA link with five $50$-GHz WDM channels.}
	\label{fig:rate_ida}
\end{figure}

The spectral efficiency lower bounds are shown in Fig.~\ref{fig:rate_ida}. For simulation, we used the split-step Fourier method for a $1000$-km optical link with IDA and standard single-mode fiber (SSMF). We used sinc pulses and the parameters in Table~\ref{tab:parameters} which are the same as those in~\cite{secondini2017fiber}. Note that the ASE noise spectral density $N_{\textrm{ASE}}=\alpha\mathcal{L}hf\eta$ is $1.13$ times smaller than in~\cite{essiambre_limits} where the phonon occupancy factor was $\eta=1.13$ rather than $\eta=1$. We used $5$ WDM channels of $50$ GHz bandwidth and $50$ GHz spacing.

We simulated a single carrier system and a multi-carrier system with $6$ subcarriers (6SC) per channel and with either uniform power allocation or FDPA with one iteration of~\eqref{eq:fdpa}. The input symbols of each subcarrier are i.i.d. circularly-symmetric Gaussian random variables. The receiver filters the center channel (consisting of one or six subcarriers) followed by digital back-propagation, matched filtering, and sampling. The subcarriers of the center channel are back-propagated jointly, and are then separated by matched filters.

We used the MPN model to compute achievable rates for the center channel (applying the WPN model with correlated noise gives similar rates). The one-sided length of the autocorrelation function $r_{\Theta}[\ell]$ is approximately $700$ symbols. A training set of $24$ sequences of $6825$ symbols ($163\:800$ total symbols) was used to estimate the parameters of the CPAN model with phase noise memory $\mu=2$ (note that $\mu \ge L-1 = 2$). A testing set of $N=120$ sequences of $6825$ symbols ($819\:000$ total symbols) was used to numerically compute the achievable rate using particle filtering and the trained CPAN model. In the 6SC systems, the one-sided length of $r_{\Theta}[\ell]$ is approximately $117$ symbols. Here, we used $24$ sequences of $1137$ symbols per subcarrier for training, and $N=120$ sequences of $1137$ symbols per subcarrier for testing. The particle filter used $K=512$ particles.

\subsection{Discussion}
Consider Fig.~\ref{fig:rate_ida}. The curve labeled ``Memoryless'' uses the phase noise model~\eqref{eq:cpan_model2} but assumes that $\{\Theta_m\}$ is i.i.d. Gaussian and $\{Z_m\}$ is circularly symmetric i.i.d. Gaussian. This is similar to the models in~\cite{essiambre_limits} and gives similar lower bounds on the spectral efficiency.

For a single carrier per channel, the CPAN model reaches a peak of $8.83$ bits/s/Hz at $-7$ dBm. This is $0.05$ bits/s/Hz larger than what the WPN model achieves in a single-carrier system~\cite{secondini2017fiber}. Moreover, the CPAN model gains 0.35 dB in power efficiency at the rate peak of the model without correlated additive noise.

For the 6SC system, the CPAN model reaches $9.01$ bits/s/Hz which slightly improves on the rates of~\cite{secondini2017fiber}. In hindsight, the small gain might have been expected given the compact autocorrelation function shown in Fig.~\ref{fig:autocorr_w}. The subcarrier approach thus compensates for the correlation even if the number of subcarriers is small.

The 6SC system with frequency-dependent power allocation (FDPA) achieves $9.09$ bits/s/Hz with the Wiener model, and $9.13$ bits/s/Hz with the CPAN model. This is a gain of $0.14$ bits/s/Hz with respect to the best bound known to us (Wiener 6SC). As can be seen in the inset, at the rate peak of Wiener 6SC, the 6SC-FDPA system with CPAN gains $0.8$ dB in power efficiency.

As already mentioned, a multi-carrier approach reduces the correlations in the additive noise which reduces the gain of the whitening filter. However, having too many subcarriers also reduces the rate~\cite{secondini2017fiber} because a narrower subcarrier bandwidth reduces the temporal correlation length which in turn reduces the NLI averaging effect  in~\eqref{eq:theta} and~\eqref{eq:nu}. The whitening filter thus provides gains even when using multiple subcarriers.

\section{Conclusion}
We used RP~\cite{mecozzi2012nonlinear} and an approximation~\eqref{eq:e-approx} to derive a CPAN model based on the NLSE. Using this model and particle filtering~\cite{dauwels2008particle, secondini2017fiber}, we computed lower bounds on the capacity of the NLSE that outperform the best bounds we have found in the literature. For example, the model gains 0.35 dB in power at the peak data rate when using a single carrier per wavelength. For six subcarriers, a FDPA scheme improves the best existing rate bounds by 0.14 bits/s/Hz and gains 0.8 dB in power at the peak data rate.

We expect that our lower bounds can be further improved by using more sophisticated receiver filters, by modeling the phase and additive noise across subcarriers as being correlated, and by shaping the symbol input distribution. Another direction for future work is extending the CPAN model to dual polarization where the correlated phase noise can likely be characterized by a $2\times 2$ Jones matrix~\cite{secondini2019nonlinearity}, and the phase and additive noise processes may become correlated across polarizations and subcarriers.

Finally, an interesting direction for future work is to build receivers that can achieve the predicted gains in real systems.

\section{Acknowledgment}
The authors wish to thank Dr. Ren\'e-Jean Essiambre for useful discussions about perturbation models. They also wish to thank a Reviewer for pointing out an error in a previous version of the paper.

\bibliographystyle{IEEEtran}
\bibliography{jlt2020.bib}

\end{document}